\documentclass[twocolumn,tighten]{aastex62}

\usepackage{graphicx}
\usepackage{amsmath}
\usepackage{comment}
\usepackage{array,multirow,color,tablefootnote}
\usepackage{xcolor}
\usepackage{subfigure}

\shorttitle{WASP-19{\rm b} Phase Curve}
\shortauthors{Wong et~al.}

\begin{document}
\title{\textit{TESS} Phase Curve of the Hot Jupiter WASP-19b} 
\correspondingauthor{Ian Wong}
\email{iwong@mit.edu}

\author[0000-0001-9665-8429]{Ian~Wong}
\altaffiliation{51 Pegasi b Fellow}
\affil{Department of Earth, Atmospheric and Planetary Sciences, Massachusetts Institute of Technology,
Cambridge, MA 02139, USA}

\author[0000-0001-5578-1498]{Bj{\" o}rn~Benneke}
\affiliation{Department of Physics and Institute for Research on Exoplanets, Universit{\' e} de Montr{\' e}al, Montr{\' e}al, QC, Canada}

\author[0000-0002-1836-3120]{Avi~Shporer}
\affil{Department of Physics and Kavli Institute for Astrophysics and Space Research, Massachusetts Institute of Technology, Cambridge, MA 02139, USA}

\author{Tara~Fetherolf}
\affiliation{Department of Physics and Astronomy, University of California, Riverside, CA 92521, USA}

\author[0000-0002-7084-0529]{Stephen~R.~Kane}
\affiliation{Department of Earth and Planetary Sciences, University of California, Riverside, CA 92521, USA}

\author[0000-0003-2058-6662]{George~R.~Ricker}
\affiliation{Department of Physics and Kavli Institute for Astrophysics and Space Research, Massachusetts Institute of Technology, Cambridge, MA 02139, USA}


\author[0000-0001-6763-6562]{Roland~Vanderspek}
\affiliation{Department of Physics and Kavli Institute for Astrophysics and Space Research, Massachusetts Institute of Technology, Cambridge, MA 02139, USA}

\author[0000-0002-6892-6948]{Sara~Seager}
\affiliation{Department of Earth, Atmospheric and Planetary Sciences, Massachusetts Institute of Technology, Cambridge, MA 02139, USA}
\affiliation{Department of Physics and Kavli Institute for Astrophysics and Space Research, Massachusetts Institute of Technology, Cambridge, MA 02139, USA}
\affiliation{Department of Aeronautics and Astronautics, Massachusetts Institute of Technology, Cambridge, MA 02139, USA}

\author[0000-0002-4265-047X]{Joshua~N.~Winn}
\affiliation{Department of Astrophysical Sciences, Princeton University, Princeton, NJ 08544, USA}



\author[0000-0001-6588-9574]{Karen~A.~Collins} 
\affiliation{Center for Astrophysics${\rm \mid}$Harvard \& Smithsonian, 60 Garden Street, Cambridge, MA 02138, USA}

\author[0000-0002-4510-2268]{Ismael~Mireles}
\affiliation{Department of Physics and Kavli Institute for Astrophysics and Space Research, Massachusetts Institute of Technology, Cambridge, MA 02139, USA}

\author{Robert~Morris}
\affiliation{NASA Ames Research Center, Moffett Field, CA 94035, USA}
\affiliation{SETI Institute, Moffett Field, CA 94035, USA}

\author{Peter~Tenenbaum}
\affiliation{NASA Ames Research Center, Moffett Field, CA 94035, USA}
\affiliation{SETI Institute, Moffett Field, CA 94035, USA}

\author[0000-0002-8219-9505]{Eric~B.~Ting}
\affiliation{NASA Ames Research Center, Moffett Field, CA 94035, USA}

\author{Stephen~Rinehart}
\affiliation{NASA Goddard Space Flight Center, Greenbelt, MD 20771, USA}

\author{Jesus~Noel~Villase{\~n}or}
\affiliation{Department of Physics and Kavli Institute for Astrophysics and Space Research, Massachusetts Institute of Technology, Cambridge, MA 02139, USA}

\begin{abstract}
We analyze the phase curve of the short-period transiting hot Jupiter system WASP-19, which was observed by the \textit{Transiting Exoplanet Survey Satellite} (\textit{TESS}) in Sector 9. WASP-19 is one of only five transiting exoplanet systems with full-orbit phase curve measurements at both optical and infrared wavelengths. We measure a secondary eclipse depth of $470^{+130}_{-110}$~ppm and detect a strong atmospheric brightness modulation signal with a semiamplitude of $319\pm51$ ppm. No significant offset is detected between the substellar point and the region of maximum brightness on the dayside. There is also no significant nightside flux detected, which is in agreement with the nightside effective blackbody temperature of $1090^{+190}_{-250}$ derived from the published \textit{Spitzer} phase curves for this planet. Placing the eclipse depth measured in the \textit{TESS} bandpass alongside the large body of previous values from the literature, we carry out the first atmospheric retrievals of WASP-19b's secondary eclipse spectrum using the SCARLET code. The retrieval analysis indicates that WASP-19b has a dayside atmosphere consistent with an isotherm at $T=2240\pm40$~K and a visible geometric albedo of $0.16\pm0.04$, indicating significant contribution from reflected starlight in the \textit{TESS} bandpass and moderately efficient day--night heat transport.

\end{abstract}

\section{Introduction}\label{sec:intro}
The \textit{Transiting Exoplanet Survey Satellite} (\textit{TESS}) Mission promises to be a watershed moment for exoplanet science. Over the course of its two-year Primary Mission, \textit{TESS} will deliver thousands of new transiting planet candidates \citep{sullivan2015,barclay2018, huang2018}. Meanwhile, it will also provide many novel avenues of study for known planetary systems. In particular, by taking advantage of the continuous, long-baseline photometry provided by \textit{TESS}, we can analyze the full-orbit visible-light phase curves of all binary systems contained within the spacecraft's coverage area.

For transiting systems, the full-orbit phase curve contains both the transit (i.e., when the planet passes in front of the host star) and the secondary eclipse (i.e., when the planet is occulted by the host star), as well as sinusoidal brightness modulations throughout the out-of-eclipse light curve. During the secondary eclipse, the light from the planet is blocked. The depth of this occultation event is a sum of the thermal emission from the planet and any reflected starlight, which is dependent on the planet's geometric albedo at the observed wavelengths.

Outside of these eclipse events, the shape of the measured phase curve is a superposition of contributions from the longitudinal variation of the planet's atmospheric brightness and photometric variability induced by gravitational interactions between the planet and the host star. For most short-period exoplanetary systems, the most prominent component in the phase curve is the atmospheric brightness modulation. These planets are expected to be tidally locked \citep[e.g.,][]{mazeh2008} and therefore have a fixed dayside hemisphere facing the star, where both thermal emission and reflected starlight are at their highest levels. Over the course of an orbit, the viewing phase of the planet varies, resulting in a periodic brightness variation with maximum near mideclipse and minimum near midtransit, i.e., a cosine of the orbital phase \citep[e.g.,][]{jenkins2003,snellen2009,shporer2015,parmentier2017}. For the most massive planets, the periodic blue- and red-shifting of the host star's spectrum and the tidal bulge raised on the star's surface can yield additional phase curve terms that are detectable in long-baseline photometry. These contributions to the total observed photometric modulation are referred to as Doppler boosting and ellipsoidal distortion; see \citet{shporer2017} for a review of these processes. 

Phase curves are a powerful tool for studying the atmospheric properties of exoplanets. Measuring the secondary eclipse depth and amplitude of the atmospheric brightness modulation provides constraints on the dayside and nightside temperatures, planetary albedo, and the efficiency of day--night heat transport. By combining the secondary eclipse depth at visible wavelengths with other measurements in the infrared (e.g., in the \textit{Spitzer} bandpasses), one can remove the degeneracy between dayside temperature and geometric albedo, and self-consistently derive robust constraints on both. An elevated geometric albedo suggests enhanced reflectivity due to the presence of clouds and/or hazes.

Here we present the \textit{TESS} phase curve of the WASP-19 system. WASP-19b is a Jupiter-sized planet ($R_{p} = 1.41~R_{J}$, $M_{p} = 1.14~M_{J}$) orbiting around an active G-dwarf with $T_{\mathrm{eff}}=5568\pm71$~K \citep{torres,mancini2013}. At 0.79~days, the orbital period of WASP-19b is the second shortest among known gas giant exoplanets \citep{hebb2010,hellier2011,tregloanreed2013}, after the recently discovered NGTS-10b \citep[0.77~days;][]{mccormac2019}. The close-in orbit and correspondingly high equilibrium dayside temperature make this planet an attractive target for atmospheric study. 

Broadband transit measurements and transmission spectroscopy observations have been carried out at both optical and infrared wavelengths spanning 0.4--5.0~$\mu$m \citep{bean2013,huitson2013,lendl2013,mancini2013,mandell2013,tregloanreed2013,sedaghati2015,sing2016,wong2016,sedaghati2017,espinoza2019}. These analyses have revealed a water vapor absorption signature at 1.4~$\mu$m consistent with roughly solar water abundance and a largely flat and featureless optical transmission spectrum \citep[e.g.,][]{iyer2016,sing2016} and a disputed detection of TiO at the day--night terminator, reported by \citet{sedaghati2017} but not seen in subsequent ground-based data analyzed by \citet{espinoza2019}. 

An extensive effort leveraging both ground- and space-based facilities has produced a plethora of secondary eclipse measurements spanning the optical and infrared, including spectroscopic eclipse measurements with the \textit{Hubble Space Telescope} and broadband observations in all four \textit{Spizter}/IRAC bands \citep{anderson2010,gibson2010,burton2012,anderson2013,abe2013,bean2013,lendl2013,zhou2013,mancini2013,zhou2014,wong2016}. In addition, full-orbit \textit{Spitzer}/IRAC phase curves at 3.6 and 4.5~$\mu$m have been published. Comparisons with both one-dimensional radiative transfer models and three-dimensional general circulation models show that the atmosphere of WASP-19b lacks a dayside temperature inversion and has relatively efficient heat transport from the dayside to the nightside when compared to other highly irradiated hot Jupiters \citep{wong2016}. With the visible-light phase curve provided by \textit{TESS}, WASP-19b becomes only the fifth exoplanet with full-orbit phase curve observations at both optical and infrared wavelengths, after HAT-P-7b \citep[e.g.,][]{borucki2009,esteves2013,wong2016}, WASP-18b \citep{maxted2013,arcangeli2019,shporer2019}, 55 Cnc e \citep[e.g.,][]{winn2011,demory2016,sulis2019}, and KELT-9b \citep{mansfield2019,wong2019kelt9}.

The paper is organized as follows. The \textit{TESS} observations and data processing are described in Section~\ref{sec:obs}. Section~\ref{sec:ana} summarizes the full phase curve model used in the light-curve fit along with our strategies for detrending long-term photometric trends due to stellar variability and instrumental systematics. The results of the phase curve analysis are presented in Section~\ref{sec:res}, and self-consistent atmospheric retrievals of the secondary eclipse spectrum are discussed in detail in Section~\ref{sec:dis}.

\section{\textit{TESS} Observations}\label{sec:obs}

\begin{figure*}
\includegraphics[width=\linewidth]{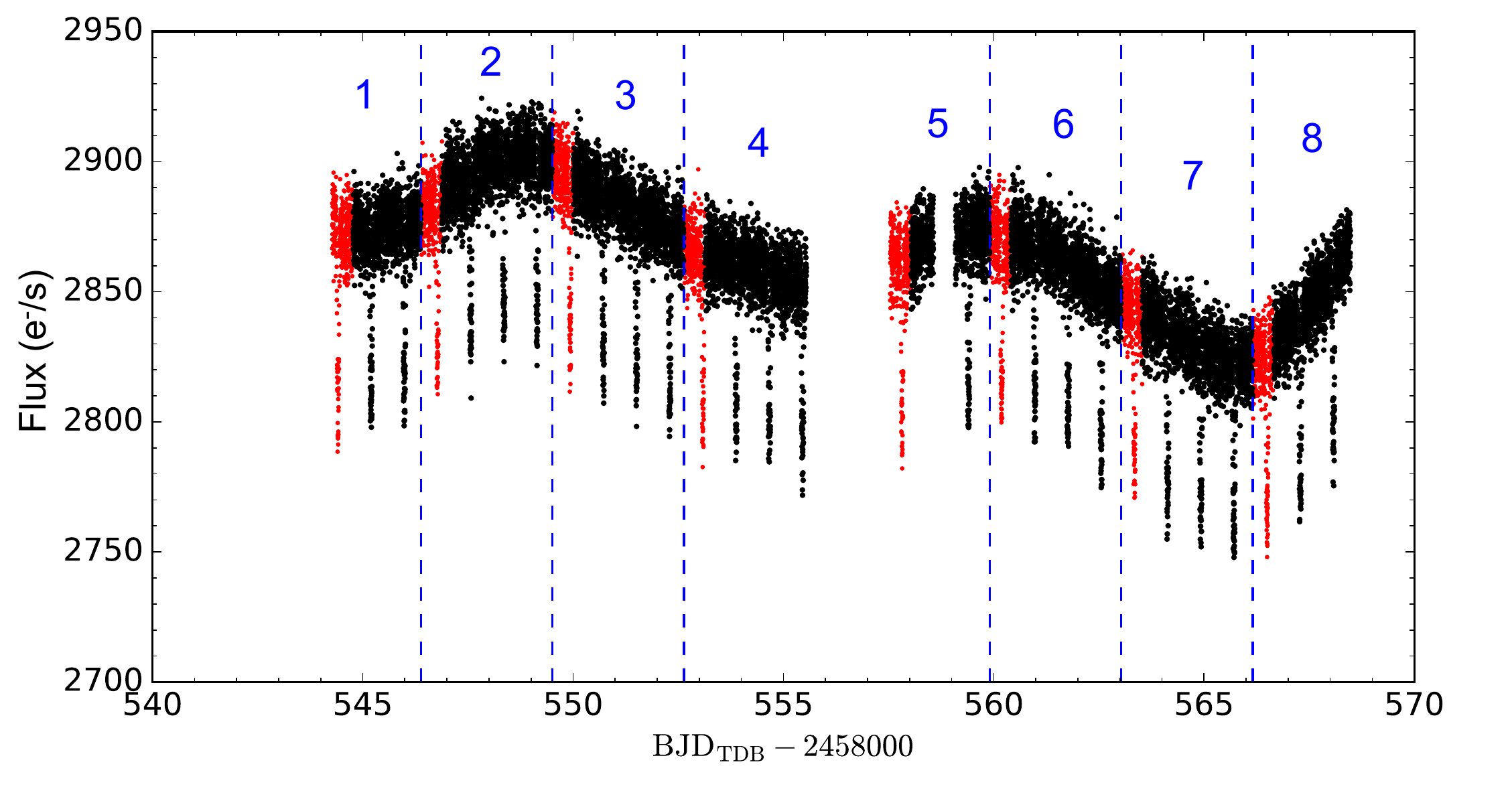}
\caption{Outlier-removed simple aperture photometry (SAP) light curve of WASP-19. The gap in the middle of time series separates the two orbits of the \textit{TESS} spacecraft during Sector 9. The vertical blue dashed lines indicate the momentum dumps. The eight segments of the light curve separated by the momentum dumps are labeled. The red points show short-term instrumental systematics and were trimmed from the time series prior to fitting. The stellar variability signal with a period of 10.5~days is evident.}
\label{fig:lc}
\end{figure*}

WASP-19 (TIC 35516889; TOI 655) was observed by Camera 2 of the \textit{TESS} spacecraft during Sector 9 (from 2019 February 28 to 2019 March 26). The downlinked data for this system consist of 11$\times$11 pixel stamps centered near the target taken at 2~minute cadence. These subarrays were passed through the Science Processing Operations Center (SPOC) pipeline \citep{jenkins2016}, which determined the optimal photometric extraction aperture and produced both Simple Aperture Photometry (SAP) and Presearch Data Conditioning (PDC) light curves. 

The PDC light curve was corrected for instrumental systematics using a linear combination of common-mode cotrending basis vectors calculated from other sources on the detector \citep{smith2012,stumpe2014}. For WASP-19, we found that the systematics detrending process introduced additional time-correlated residual features in the photometry that are not present in the SAP light curve. Therefore, in this analysis, we used the cleaner SAP light curve.

We first removed all points with a nonzero quality flag value before applying a 16~point-wide moving median filter to the eclipse-masked light curve to remove $3\sigma$ outliers. The resultant light curve is shown in Figure~\ref{fig:lc}. The gap in the middle of the time series divides the Sector into two spacecraft orbits and indicates when the observations were paused to downlink the on-board data to Earth. WASP-19 is an active star, and the primary long-term variation (at the $\sim$3\% level) in the light curve can be attributed to stellar variability, which has a measured period of roughly 10.5~days \citep[e.g.,][]{hebb2010,espinoza2019}.

During nominal \textit{TESS} spacecraft operations, momentum dumps are scheduled three times per orbit, when the thrusters are fired to reduce the speed of the on-board reaction wheels and decrease the pointing jitter. In Sector 9, these occurred every 3.12~days (see Figure~\ref{fig:lc}). In addition to small discontinuities in the photometry during each event, there are typically short-term flux variations in the vicinity of the momentum dumps, which can affect the modeling of time-dependent astrophysical signals. These instrumental signals occur on timescales that are comparable to those of the astrophysical phase curve variation ($\sim$1~day), so strategies to mitigate these features, such as polynomial detrending, can introduce systematic biases to and/or spurious correlations with the astrophysical parameters of interest. For the WASP-19 system, the potential for these residuals to produce significant biases is particularly severe, because the interval between momentum dumps is almost exactly four times the orbital period ($P=0.79$~day), meaning that the events always occur near the same orbital phase.

To evaluate the extent of instrumental flux variations around the momentum dumps, we split the light curve into eight segments separated by the momentum dumps. We then fit each segment individually to the combined phase curve and stellar variability model described in Sections~\ref{subsec:model} and \ref{subsec:var}, and inspected the resultant residual series binned in 5--10~minute intervals to make time-correlated noise more apparent. We found that many of the segments showed clear short-term flux ramps and/or increased scatter within the $\sim$0.5~days of photometry immediately following a momentum dump or the start of a spacecraft orbit. The flux ramps in particular strongly biased the best-fit phase curve amplitudes, producing unphysically large amplitudes at the fundamental of the cosine of the orbital period (corresponding to the atmospheric brightness modulation). Therefore, we chose to trim the first 0.5~days of data from each segment. Trimming anywhere between 0.25 and 1.0~day of data did not yield any significant change to the fitted parameters. Overall, outlier removal and data trimming removed 19.4\% of points from the raw SAP light curves. The trimmed points are denoted in red in Figure~\ref{fig:lc}. 

Given the relatively large pixel scale of $21''$, a target's pixel response function (PRF) may overlap with those of neighboring stars. For each light curve, the SPOC pipeline provides an estimate of the level of contamination contained within the extraction aperture via the CROWDSAP value stored in the header of the light-curve files. For WASP-19, the CROWDSAP value is 0.8927, which indicates that 10.73\% of the flux extracted from the optimal aperture is contaminated by nearby sources. Prior to normalizing the light curves, we corrected for this flux dilution by subtracting 10.73\% of the overall median flux from the photometric series. 

The level of contamination in the extraction aperture from nearby stars is calculated using a model PRF derived from commissioning data. The accuracy of the deblending depends on the uncertainty in the PRF model as well as intrinsic variations in the PRFs of different sources across the detector; the relative uncertainty in the computed CROWDSAP value is estimated to be at the level of a few percent \citep{jenkins2010}. It follows that the relative contribution of possible detrending uncertainty to the measured astrophysical parameters is at most a few tenths of a percent. This is significantly smaller than any of the relative uncertainties from our joint fits (see Section~\ref{sec:res}), and we can therefore neglect this error contribution in our analysis.

\section{Data Analysis}\label{sec:ana}
The phase curve modeling in this work is similar to the previously published analyses of the WASP-18b and KELT-9b phase curves \citep{shporer2019,wong2019kelt9}. We utilized the transit, eclipse, and phase curve fitting pipeline ExoTEP to carry out the data extraction and analysis. Detailed technical descriptions of the data analysis methods in ExoTEP can be found in \citet{benneke2019} and \citet{wong2019}.

\begin{deluxetable*}{lllll}
\tablewidth{0pc}
\tabletypesize{\scriptsize}
\tablecaption{
    Results of Joint Fits
    \label{tab:fit}
}
\tablehead{\\ & \multicolumn{2}{c}{\underline{Fit A\tablenotemark{a}}} & \multicolumn{2}{c}{\underline{Fit B\tablenotemark{a}}} \\
    \colhead{Parameter} &
    \colhead{Value}                     &
    \colhead{Error}  &
    \colhead{Value}                     &
    \colhead{Error}    
}
\startdata
\sidehead{\textit{Fitted Parameters}}
$R_p/R_*$     & 0.15243 & $_{-0.00086}^{+0.00085}$  & 0.15244 & $_{-0.00089}^{+0.00085}$ \\
$T_0$ (BJD$_{\mathrm{TDB}}-2458000$)        & 555.45470 & $_{-0.00011}^{+0.00010}$ & 555.45472 & $_{-0.00011}^{+0.00010}$\\
$P$ (days)    & 0.788849 & $_{-0.000011}^{+0.000010}$ & 0.788846 & $_{-0.000010}^{+0.000011}$\\
$b$           & 0.637 & $_{-0.016}^{+0.015}$ & 0.638 & $_{-0.018}^{+0.016}$\\
$a/R_*$       & 3.612 & $_{-0.056}^{+0.057}$ & 3.606 & $_{-0.057}^{+0.062}$\\
$\bar{f_p}$ (ppm)    & 163 & $_{-97}^{+113}$ & 163 & $_{-99}^{+110}$ \\
$A_1$ (ppm)   & 49  & $_{-44}^{+45}$ &$\dots$&$\dots$\\
$A_2$ (ppm)   & $-$86   & 38 &$\dots$&$\dots$\\
$B_1$ (ppm)   & $-$313 & 50 & $-$319 & $_{-50}^{+52}$ \\
$B_2$ (ppm)   & 12 & $_{-48}^{+51}$  &$\dots$&$\dots$\\
$\delta$ ($^{\circ}$)    &$\dots$&$\dots$& 9.7 & $_{-6.9}^{+7.4}$ \\
$\sigma_1$ (ppm)   & 3046 & 66 & 3051 & $_{-59}^{+63}$ \\
$\sigma_2$ (ppm)   & 3319 & $_{-54}^{+55}$ & 3324 & $_{-53}^{+55}$\\
$\sigma_3$ (ppm)   & 3173 & $_{-52}^{+53}$ & 3172 & $_{-55}^{+52}$\\
$\sigma_4$ (ppm)   & 3179 & $_{-51}^{+55}$ & 3183 & 52\\
$\sigma_5$ (ppm)   & 3337 & $_{-76}^{+79}$ & 3328 & $_{-75}^{+78}$\\
$\sigma_6$ (ppm)   & 3177 & $_{-54}^{+53}$ & 3172 & $_{-50}^{+51}$\\
$\sigma_7$ (ppm)   & 3239 & $_{-50}^{+56}$ & 3228 & $_{-54}^{+51}$\\
$\sigma_8$ (ppm)   & 3217 & $_{-66}^{+65}$ & 3219 & 61\\
\sidehead{\textit{Derived Parameters}} 
Transit depth (ppm)\tablenotemark{b}    & 23240 & 260 & 23240 & $_{-270}^{+250}$\\
$i$ ($^{\circ}$)      & 79.84 & 0.41 & 79.80 & $_{-0.42}^{+0.46}$ \\
Secondary eclipse depth, $D_{d}$ (ppm)  & 477 & $_{-104}^{+120}$ & 473 & $_{-106}^{+131}$\\
Nightside flux, $D_{n}$ (ppm)  & $-$150 & $_{-110}^{+130}$ & $-$140 & 110
\enddata
\textbf{Notes.}
\tablenotetext{a}{Fit A allowed all four Fourier coefficients ($A_1$, $A_2$, $B_1$, $B_2$) to vary, while fixing the phase shift of the atmospheric brightness modulation $\delta$ to zero. Fit B fixed all phase curve parameters to zero, except for the semiamplitude and phase shift of the atmospheric brightness modulation ($B_1$ and $\delta$).}
\tablenotetext{b}{Calculated as $(R_2/R_1)^2$.}
\end{deluxetable*}

\subsection{Full phase curve model}\label{subsec:model}
The astrophysical phase curve variation is described by a harmonic series that contains both the fundamental and first harmonic of the cosine and sine as a function of the orbital phase $\phi(t)\equiv 2\pi(t-T_0)/P$, where $T_0$ is the midtransit time, and $P$ is the orbital period. The primary time-dependent signal in the planet's flux $\psi_{p}(t)$ is the atmospheric brightness modulation, which varies at the fundamental of the cosine, with an amplitude $B_1$; since the planet's observed brightness is maximum near secondary eclipse and minimum near midtransit, the expected value of $B_{1}$ is negative. The other harmonics are assigned to the host star's flux variation $\psi_{*}(t)$ and include the characteristic terms corresponding to Doppler boosting ($A_1$: fundamental of sine; \citealt{shakura1987,loeb2003,zucker2007}) and the leading term of the ellipsoidal distortion signal ($B_2$: first harmonic of cosine; \citealt{morris1985,morris1993,pfahl2008}); a nonzero $A_{2}$ amplitude is not expected in the nominal case, but may arise due to, for example, a phase shift in the ellipsoidal distortion signal, as was detected in the \textit{TESS} phase curve of KELT-9b \citep{wong2019kelt9}:
\begin{align}
\label{planet}\psi_{p}(t) &= \bar{f_{p}} + B_1 \cos(\phi+\delta),\\
\label{star}\psi_{*}(t) &= 1+ A_1 \sin(\phi) +  A_2 \sin(2\phi) +  B_2 \cos(2\phi).
\end{align}
Here, $\bar{f_{p}}$ is a free parameter that represents the mean brightness of the planet relative to the host star, and $\delta$ represents the phase shift of the atmospheric brightness modulation. Such a phase shift can be interpreted as an offset of the hottest region of the dayside hemisphere relative to the substellar point, or an asymmetric reflected light distribution on the dayside due to nonuniform cloud coverage (or a combination of the two effects). Given the sign convention used in Equation~\eqref{planet}, a positive value of $\delta$ indicates that the maximum observer-facing hemispheric brightness occurs before secondary eclipse, corresponding to an eastward shift of the dayside hotspot. From here, the secondary eclipse depth (dayside brightness at superior conjunction) and nightside flux are derived as follows: $D_{d}=\bar{f_{p}}+B_1\cos(\pi+\delta)$ and $D_{n}=\bar{f_{p}}+B_1\cos(\delta)$.

Both transits and secondary eclipses are modeled in ExoTEP using BATMAN \citep{kreidberg2015}. We fit for the planet--star radius ratio $R_{p}/R_{*}$, transit ephemeris (midtransit time $T_{0}$, orbital period $P$), and transit shape (impact parameter $b$, scaled semi-major axis $a/R_{*}$). For the midtransit time, we designated the zeroth epoch to be the transit event closest to the median of the combined time series. The orbit of WASP-19b is circular to within the uncertainties
 \citep[$e<0.02$ at $3\sigma$;][]{hebb2010,hellier2011}, so we fixed eccentricity to zero.

The measured stellar parameters of WASP-19 are $T_{\mathrm{eff}}=5568\pm71$~K, $\log g=4.45\pm0.05$, and $\mathrm{[Fe/H]}=0.15\pm0.07$ \citep{torres}. We fixed the quadratic limb-darkening coefficients to the values computed in \citet{claret2018} for the nearest combination of stellar parameters ($T_{\mathrm{eff}}=5600$~K, $\log g=4.50$, and $\mathrm{[Fe/H]}=0.0$): $u_{1}=0.3799$ and $u_2=0.2051$. We find similar quadratic limb darkening coefficients to those derived from analyses of previously published broadband transit measurements at wavelengths spanning the \textit{TESS} bandpass. For example, \citet{sedaghati2015} derived $u'_{1}=0.391\pm0.094$ and $u'_{2}=0.225\pm0.051$ for a broadband transit light curve spanning 550--830~nm. We also experimented with fitting for the limb-darkening coefficients in the joint fit, but found that the intrinsic precision of the photometry is too poor to produce reasonable constraints. Moreover, when allowing the limb-darkening coefficients to vary freely, we obtained astrophysical parameter measurements that agree with the values from the fits with fixed limb darkening coefficients to well within $1\sigma$, albeit with larger uncertainties.

Combining the out-of-eclipse phase curve variation in Equations~\eqref{planet} and \eqref{star} with the transit and eclipse light-curve models $\lambda_t(t)$ and $\lambda_e(t)$, we obtain the full phase curve model, with the mean uneclipsed flux normalized to unity:
\begin{equation}
    \psi(t) = \frac{\lambda_t(t)\psi_{*}(t)+\lambda_e(t)\psi_{p}(t)}{1+\bar{f_p}}.
\end{equation}

\subsection{Stellar variability}\label{subsec:var}
As demonstrated in Figure~\ref{fig:lc}, the most salient time-dependent flux variation in the light curve apart from the orbital phase curve is stellar variability. In our analysis, we defined a detrending model for each segment to capture both the host star's long-term rotational flux variation and any instrumental systematics in the SAP light curve using a generalized low-order polynomial in time
\begin{equation}\label{systematics}
    S_N^{\lbrace i\rbrace}(t) = \sum\limits_{j=0}^{N}c_j^{\lbrace i\rbrace}(t-t_0)^j,
\end{equation}
where $t_0$ is the timestamp of the first exposure in the segment $i$, and $N$ is the order of the detrending polynomial. The complete flux model used in our fits is 
\begin{equation}
    f(t) = S_N^{\lbrace i\rbrace}(t)|_{i=1-8}\times\psi(t).
\end{equation}

We determined the optimal polynomial orders by carrying out individual fits of each segment and selected the order that minimized the Bayesian Information Criterion (BIC): $\mathrm{BIC}\equiv k\log n -2 \log L$, where $k$ is the number of free parameters in the fit, $n$ is the number of data points in the segment, and $L$ is the maximum log-likelihood. For segments 1--8, the polynomial orders we used were 3, 2, 1, 1, 1, 3, 3, and 2, respectively. When altering the polynomial orders by $\pm$1--2, we found that the resulting astrophysical quantities of interest (secondary eclipse depth and phase curve amplitudes) from the joint fit did not vary by more than $0.3\sigma$.

\subsection{Model fit}\label{subsec:fit}

\begin{figure*}
\includegraphics[width=\linewidth]{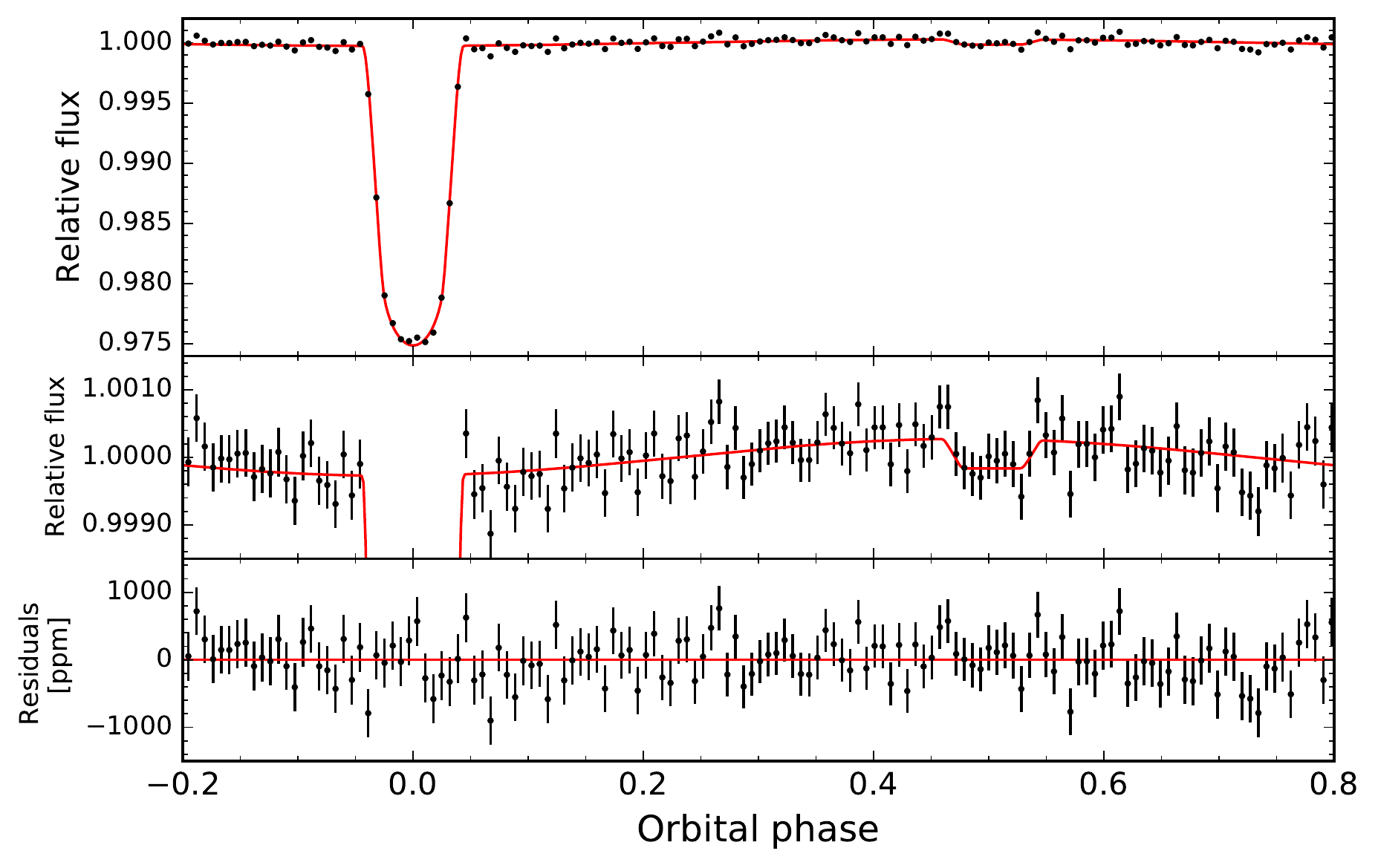}
\caption{Top panel: phase-folded light curve of WASP-19 after correcting for stellar variability and long-term trends, binned in 8~minute intervals (black points), along with the best-fit full phase curve model from our joint analysis (red line). The phase curve model shown is from Fit B, in which only the amplitude ($B_1$) and phase shift ($\delta$) of the atmospheric brightness modulation were allowed to vary. Middle panel: same as top panel, with a stretched vertical axis to detail the phase curve modulation and secondary eclipse. Bottom panel: corresponding residuals from the best-fit model.}
\label{fig:fit}
\end{figure*}

We carried out two separate joint fits of all eight segments in the WASP-19 light curve. In the first fit (referred to hereafter as Fit A), the free parameters are $R_p/R_*$, $T_0$, $P$, $b$, $a/R_*$, $\bar{f_p}$, $A_1$, $A_2$, $B_1$, $B_2$, and $c_N^{\lbrace i\rbrace}$. Including both the fundamental of the sine and cosine makes the phase shift parameter $\delta$ degenerate, so we fixed $\delta=0$ in this fit. In order to ensure realistic uncertainties on the astrophysical parameters given the intrinsic scatter of the data points, we fit for a uniform per-point uncertainty $\sigma_i$ for each segment such that the resultant reduced chi-squared value is unity. The total number of free astrophysical, detrending, and noise parameters for this fit is 42. For the second joint fit (Fit B), we focused on the planet's atmospheric brightness modulation and allowed only $B_1$ and $\delta$ to vary freely, while fixing the other phase curve amplitudes ($A_1$, $A_2$, and $B_2$) to zero. The total number of free parameters in Fit B is 40.

In ExoTEP, the best-fit values and posterior distributions of all free parameters are computed simultaneously within the affine-invariant Markov Chain Monte Carlo (MCMC) framework \texttt{emcee} \citep{emcee}. We set the number of walkers to four times the number of free parameters and initiated each chain near the best-fit parameter values from the individual segment fits. The length of each chain was 25,000 steps, and we discarded the first 60\% of each chain before calculating the posterior distributions. As a test for convergence, we ran the Gelman--Rubin test \citep{gelmanrubin} and ensured that the diagnostic value $\hat{R}$ was below 1.1. In addition, we ran a series of joint fits using the same chain length and compared the parameter estimates to ensure that they were self-consistent to within $0.1\sigma$.

\section{Results}\label{sec:res}
The median and $1\sigma$ uncertainties of all astrophysical and noise parameters from the two joint fits are listed in Table~\ref{tab:fit}. Comparing the BIC values for the two joint fits, we obtain $\Delta\mathrm{BIC}=17.1$ in favor of Fit B. Figure~\ref{fig:fit} shows the combined phase-folded light curve with long-term trends due to stellar variability and instrumental systematics removed, along with the best-fit full phase curve model from Fit B. Both the secondary eclipse and the out-of-eclipse phase curve modulation are clearly discernible.

From the results for Fit A, the only phase curve amplitude measured at high statistical significance is $B_1$ --- the fundamental of the cosine at the orbital period, corresponding to the atmospheric brightness modulation --- with the other three values consistent with zero to within $2.3\sigma$. Using literature values for the stellar and planetary masses and orbital parameters, we estimated the expected amplitude of Doppler boosting ($A_1$) and ellipsoidal distortion ($B_2$) to be roughly 4 and 30~ppm, respectively \citep[e.g.,][]{esteves2013}; these values are smaller than or comparable to the error bars on the corresponding phase curve amplitudes, while being formally consistent with the fitted parameter values computed in the light-curve analysis. Meanwhile, we measured a weak signal at the first harmonic of the sine with an amplitude of $|A_2|=86\pm38$~ppm. Given the small predicted ellipsoidal distortion amplitude, this unexpected signal is not likely to be a phase shift in the ellipsoidal distortion modulation, but rather likely attributable to uncorrected systematics in the light curve.

Fit B allowed only the atmospheric modulation amplitude and its phase shift to vary, yielding $B_1=-319^{+52}_{-50}$~ppm, a $6.1\sigma$ detection consistent with the value from Fit A. Meanwhile, we measured a small eastward phase shift in the atmospheric brightness signal of roughly $10^{\circ}$, though the estimate is consistent with zero at $1.4\sigma$. The analysis of \textit{Spitzer}/IRAC phase curves by \citet{wong2016} detected eastward shifts in the dayside thermal emission maximum of $10.5\pm4.0$ and $12\overset{\circ}{.}9\pm3\overset{\circ}{.}6$ at 3.6 and 4.5~$\mu$m, respectively. These more statistically robust measurements are consistent with the phase shift derived from the \textit{TESS} light curve and suggest the presence of superrotating equatorial winds. For all other astrophysical parameters, the values from Fits A and B are consistent with each other to well within $1\sigma$. We utilize the parameter estimates from Fit B in the subsequent discussion.

Combining the average planetary brightness and the semiamplitude of the atmospheric brightness modulation, we derived a secondary eclipse depth of $473^{+131}_{-106}$~ppm ($4.5\sigma$). Meanwhile, the calculated median nightside flux is negative ($D_{n}=\bar{f_{p}}+B_1\cos(\delta) = -140\pm110$~ppm; Section~\ref{subsec:model}), while consistent with zero at $1.3\sigma$, indicating negligible thermal emission in the \textit{TESS} bandpass. The nightside emission spectrum derived from \textit{Spitzer}/IRAC phase curves matches a single blackbody with an effective temperature of $1090^{+190}_{-250}$~K \citep{wong2016}; extrapolating to \textit{TESS} wavelengths yields a predicted $3\sigma$ upper limit on the nightside flux of 12~ppm.

The orbital period and midtransit time computed by our joint fits are consistent with the latest published transit ephemerides at better than $1\sigma$ \citep{espinoza2019}. Meanwhile, given the large number of transits contained within the \textit{TESS} light curve, we obtained precise best-fit values for scaled semi-major axis and orbital inclination ($a/R_{*}=3.606^{+0.062}_{-0.057}$, $i=79.80^{+0.46}_{-0.42}$~deg) that are generally consistent with previous estimates \citep[e.g.,][]{hebb2010,mancini2013,tregloanreed2013,sedaghati2017,espinoza2019}.

Turning to the transit depth, we find that our value ($23,240^{+250}_{-270}$~ppm) is significantly larger than other measurements in the literature. \citet{hebb2010} lists $20,300\pm400$~ppm, while other visible-light transit measurements have yielded depths spanning 19,000--21,000~ppm \citep{lendl2013,mancini2013,tregloanreed2013,sedaghati2015,sedaghati2017,espinoza2019}. Our measured transit depth is roughly 10--15\% deeper than previous measurements. We note that WASP-19 is a highly active G-dwarf, and previous photometric modeling of the star during 2017 showed brightness modulation of up to 5\% across the characteristic $\sim$10.4~day period of stellar variability \citep{hebb2010}, with long-term changes to the star's average flux of a few percent \citep{espinoza2019}. In this context, the deeper transit depth we measured suggests that the stellar surface was significantly more spotty during the \textit{TESS} observations than during previous epochs; however, there was no contemporaneous photometric monitoring of WASP-19 during 2019. We inspected each individual transit light curve for signs of spot-crossing events, as have been seen in some previous transit observations \citep[e.g.,][]{mancini2013,tregloanreed2013,sedaghati2015,espinoza2019}, but we did not find any such events to the level of precision in the data. 

\textit{TESS} will revisit the Southern Sectors during the first year of the Extended Mission. These upcoming observations of the WASP-19 system will yield an independent transit depth measurement and allow us to quantitatively assess variations in the brightness and activity level of WASP-19 on year-long timescales.

The median scatter across the full Sector 9 light curve, binned by 30~minute and 60~minute intervals, is 840 and 600~ppm, respectively. Meanwhile, the corresponding phase-folded median scatter values are 180 and 130~ppm.

\section{Discussion}\label{sec:dis}

\begin{deluxetable}{cccc}[t!]
\tablewidth{0pc}
\tabletypesize{\scriptsize}
\tablecaption{
    WASP-19b Secondary Eclipses
    \label{tab:eclipses}
}
\tablehead{
    \colhead{$\lambda$ ($\mu$m)\tablenotemark{a}} &
    \colhead{$\Delta\lambda$ ($\mu$m)\tablenotemark{a}} &
    \colhead{Depth (ppm)}                     &
    \colhead{Reference}  
}
\startdata
0.685 & 0.265 & $390\pm190$ &  \citet{abe2013}\\
0.7865 & 0.2000 & 473$_{-106}^{+131}$ & This work \\
0.79779 & 0.14288 & $480\pm130$ & \citet{mancini2013}\\
0.866 & 0.052 & $800 \pm 290$ & \citet{zhou2013}\\
0.9665 & 0.1279 & $352\pm116$ & \citet{lendl2013}\\
0.9665 & 0.1279 & $880\pm190$ & \citet{burton2012}\\
1.186 & 0.006 & $1711_{-726}^{+745}$ & \citet{lendl2013}\\
1.325 & 0.075 & $830\pm 390$ & \citet{bean2013}\\
1.45 & 0.05 & $2080\pm450$ & \citet{bean2013}\\
1.55 & 0.05 & $1800\pm170$ & \citet{bean2013}\\
1.62 & 0.289 & $2590_{-440}^{+460}$ & \citet{anderson2010}\\
1.65 & 0.05 & $2000\pm360$ & \citet{bean2013}\\
1.825 & 0.125 & $1880\pm380$ & \citet{bean2013}\\
2.00 & 0.05 & $2380\pm300$ & \citet{bean2013}\\
2.095 & 0.010 & $3660\pm720$ & \citet{gibson2010}\\
2.10 & 0.05 & $2270\pm160$ & \citet{bean2013}\\
2.144 & 0.162 & $2870\pm200$ & \citet{zhou2014}\\
2.20 & 0.05 & $2420\pm310$ & \citet{bean2013}\\
2.30 & 0.05 & $3120\pm910$ & \citet{bean2013}\\
3.6 & 0.375 & $4850\pm240$ & \citet{wong2016}\\
4.5 & 0.5075 & $5840\pm290$ & \citet{wong2016}\\
5.8 & 0.7 & $6500\pm1100$ & \citet{anderson2013}\\
8.0 & 1.45 & $7300\pm1200$ & \citet{anderson2013}
\enddata
\textbf{Note.}
\tablenotetext{a}{Center and half-width of bandpass, in microns.}
\end{deluxetable}

\begin{figure*}
\includegraphics[width=\linewidth]{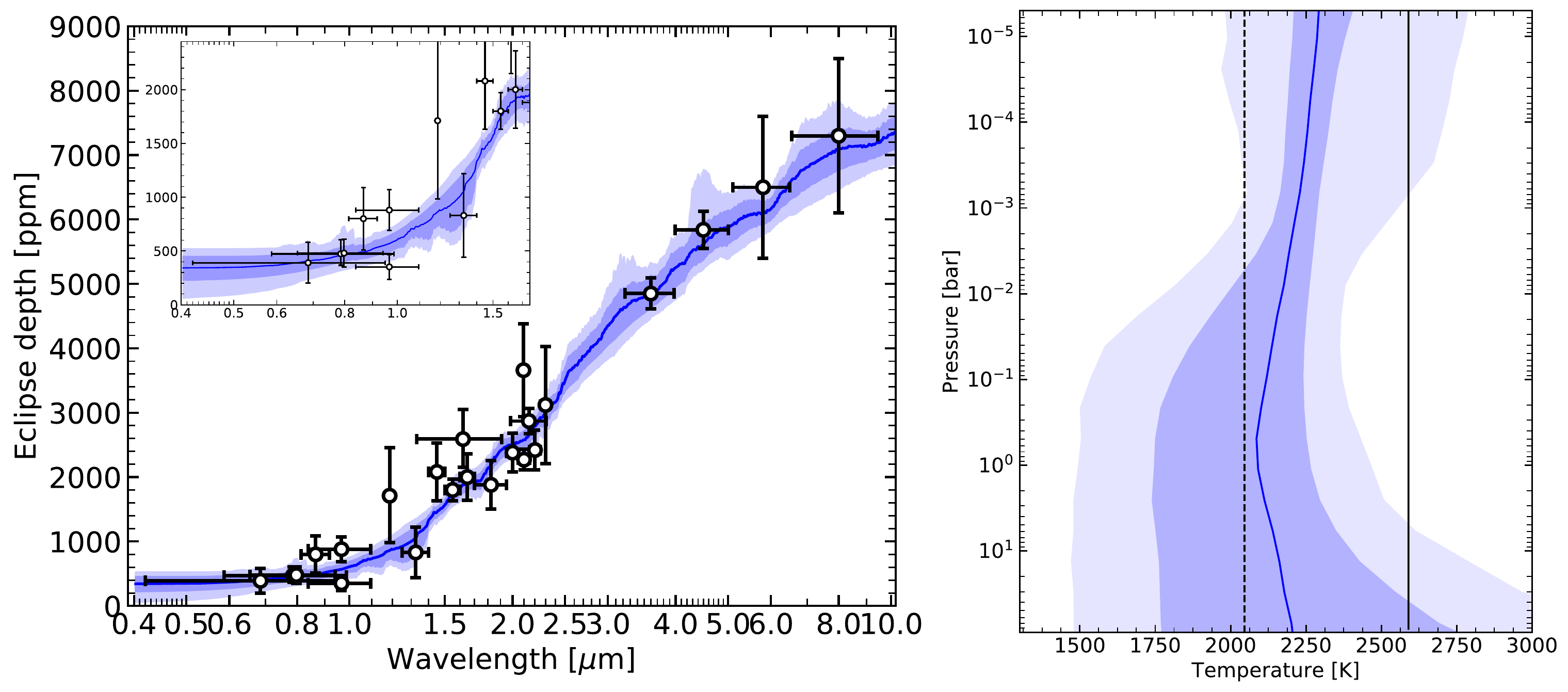}
\caption{Left: secondary eclipse spectrum of WASP-19b. The published observations and are shown in black. The solid blue line denotes the median atmospheric model, while the shaded regions indicate $1\sigma$ and $2\sigma$ bounds. The infrared data are well-matched by near-isothermal atmospheres with $T\sim2200$~K. The inset plot is an expanded view of the optical wavelength region. Measurements at wavelengths shorter than about 1~$\mu$m require a significant contribution of reflected light. Right: the retrieved $T$--$P$ profile; the median curve, $1\sigma$, and $2\sigma$ bounds are shown in blue. The atmosphere is consistent with an isotherm at $T\sim2200$~K. The vertical black lines indicate the two limiting cases for the dayside equilibrium temperature, assuming a Bond albedo of 0.1: homogeneous heat redistribution (dashed) and instant reradiation (solid).}
\label{fig:retrieval}
\end{figure*}

The ultra-short orbital period and high levels of incident stellar irradiation have made WASP-19b one of the most well-characterized exoplanets in the literature. Combining the \textit{TESS} phase curve presented in this work with the full-orbit \textit{Spitzer}/IRAC phase curves at 3.6 and 4.5~$\mu$m \citep{wong2016} provides a longitudinally resolved broadband optical--infrared emission and reflectance spectrum of the planet. Our measurement of the secondary eclipse depth in the \textit{TESS} bandpass adds to the large body of previously published eclipse depths spanning almost the entire wavelength range between 0.6 and 10~$\mu$m; the full list of literature values is compiled in Table~\ref{tab:eclipses}. Particularly notable is the earlier ground-based $i'$-band eclipse measurement of $480\pm130$~ppm \citep{mancini2013}, which is consistent with our value to within $0.1\sigma$.

Leveraging the full range of secondary eclipse measurements available for the WASP-19 system, we carried out the first retrieval analysis of the secondary eclipse spectrum in order to place directly constrain the dayside temperature--pressure ($T$--$P$) profile and geometric albedo of WASP-19b.

We utilized the SCARLET atmospheric retrieval code \citep[e.g.,][]{bennekeseager1,bennekeseager2,benneke2015,benneke2019}, which calculates the posterior distributions of all free parameters simultaneously within a Bayesian MCMC framework (see the references for a full description of the modeling). We ran two retrievals: (1) a free chemical retrieval, in which we allowed the mixing ratios of six molecular gases relevant in hot gaseous exoplanet atmospheres --- H$_2$O, CH$_4$, CO, CO$_2$, NH$_3$, and HCN --- to vary freely within a background atmosphere of hydrogen and helium (in solar abundances), and (2) a simple isothermal retrieval, where we fix the atmospheric composition to solar and assume a single dayside temperature.

In order to constrain the temperature--pressure profile of the dayside atmosphere in the free chemical retrieval, SCARLET incorporates the five-parameter analytical formulation described in \citet{parmentier2014}. We have augmented this simple model by ensuring that the retrieved $T$--$P$ profiles are physically plausible, i.e., that the wavelength-integrated thermal emission at each step is consistent with the incident stellar irradiation, a Bond albedo between 0 and 0.7, and day--night heat redistribution values between full heat redistribution across the planet and no heat redistribution. In both retrieval runs, the contribution of reflected starlight is accounted for by the inclusion of a wavelength-independent geometric albedo $A_{g}$. For each model atmosphere generated in the retrievals, SCARLET produces a high-resolution emission spectrum using line-by-line radiative transfer and then integrates the spectrum over the respective instrument response functions to generate the synthetic spectrum for comparison with the data. 

The results of our free chemical retrieval run are shown in Figure~\ref{fig:retrieval}. The median emission+reflectance spectrum model is plotted, along with $1\sigma$ and $2\sigma$ bounds; at all wavelengths, the observed secondary eclipse depths are well-matched by the model to better than $1.5\sigma$ in the measurement uncertainties. The $T$--$P$ profile does not show any notable structure and is consistent with an isotherm that agrees with the blackbody brightness temperature of $2372\pm60$~K derived from the measured \textit{Spitzer} 3.6 and 4.5~$\mu$m secondary eclipses alone \citep{wong2016}. Meanwhile, no constraints on the molecular gas abundances were obtained from the data.

From the free chemical retrieval, we obtained a geometric albedo of $0.20\pm0.07$. The inset plot in Figure~\ref{fig:retrieval} shows that the relatively high-precision \textit{TESS}-band secondary eclipse measurement from this work and the previous $i'$-band eclipse depth from \citet{mancini2013} are instrumental in solidifying the presence of nonnegligible reflected light at visible wavelengths. This measurement is consistent with the 97.5\% confidence upper limit of 0.21 derived by \citet{mallonn2019} based on the $z'$-band eclipse detection in \citet{lendl2013}. 

\begin{figure}
\includegraphics[width=\linewidth]{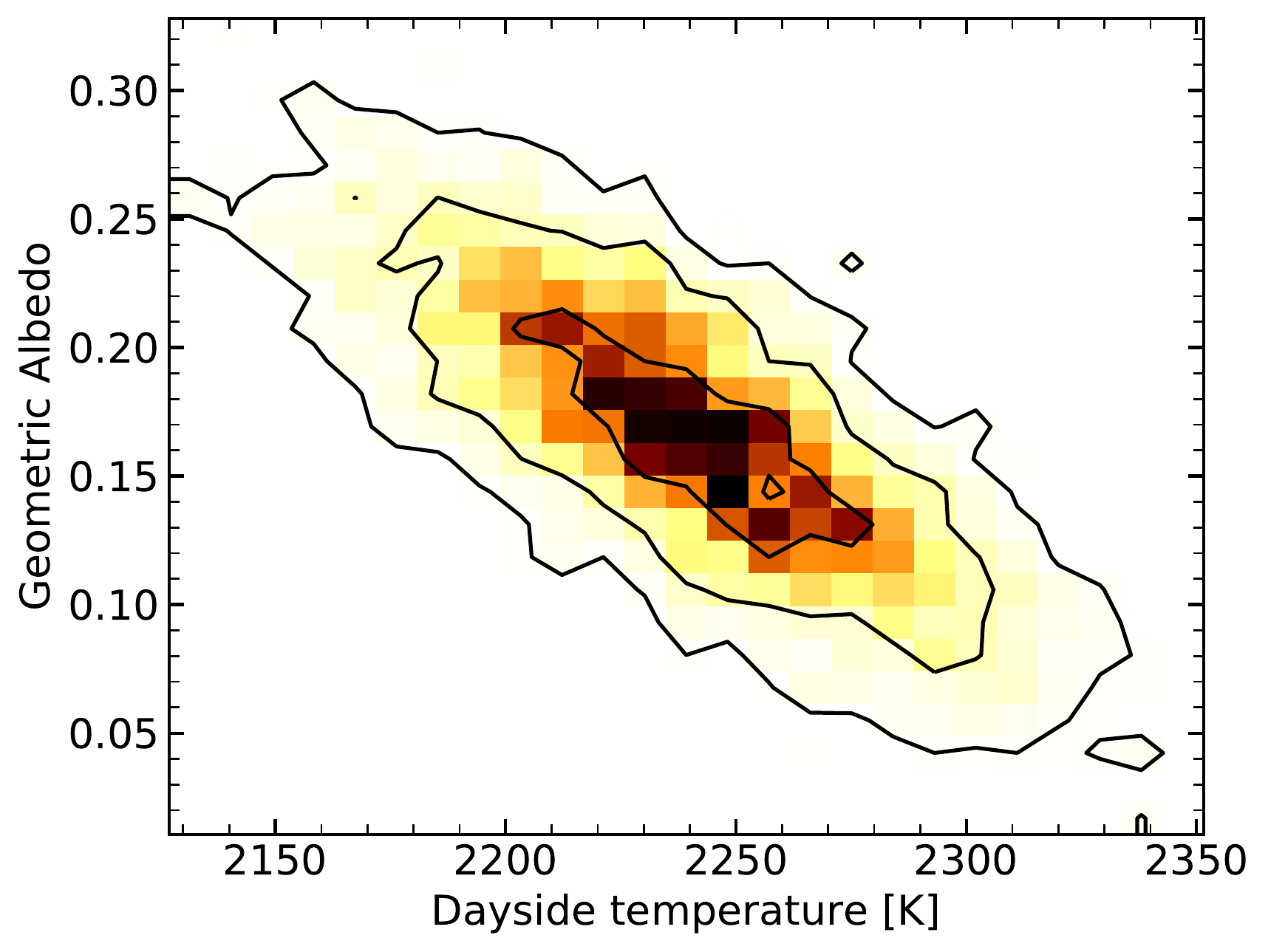}
\caption{Marginalized two-dimensional posterior distribution of dayside temperature and optical geometric albedo from the isothermal retrieval run. The black contours denote 1, 2, and $3\sigma$ bounds. The relatively high-precision secondary eclipse measurements at both visible and thermal infrared wavelengths allow for tight constraints on both quantities: $T_{\mathrm{day}}=2240\pm40$~K and $A_{g}=0.16\pm0.04$.}
\label{fig:isotherm}
\end{figure}

We note that the unconstrained molecular composition and temperature--pressure structure in our free chemical retrieval mean that the corresponding constraint on the albedo is maximally conservative. The isothermal retrieval run provides stronger constraints on both dayside temperature and geometric albedo. Figure~\ref{fig:isotherm} shows the two-dimensional posterior distribution of these quantities that we obtained from the isothermal atmospheric retrieval: $T_{\mathrm{day}}=2240\pm40$~K and $A_{g}=0.16\pm0.04$. These results agree with the less constrained values from the free chemical retrieval. 

From our retrieval analysis of the full secondary eclipse spectrum, we have produced the first direct measurement of WASP-19b's optical geometric albedo. The measured albedo is significantly nonzero ($3\sigma$--$4\sigma$) and is broadly consistent with the range of literature values for other short-period gas giants \citep[see, for example, the albedos derived from \textit{Kepler} secondary eclipses;][]{hengdemory,angerhausen2015,esteves2015}. We note that the majority of published optical geometric albedos are not direct constraints, but rather values derived from single-band visible-wavelength secondary eclipses given particular assumptions on the dayside brightness temperature and/or day--night heat redistribution. 

Only a handful of hot Jupiters have direct albedo measurements, which require robust secondary eclipse detections at both visible and thermal infrared wavelengths in order to break the inherent degeneracy between dayside temperature and geometric albedo. Planets with direct optical geometric albedo measurements or upper limits include HD 189733b ($<$0.12 across 450--570~nm; \citealt{evans2013}), HD 209458b ($0.038\pm0.045$; \citealt{rowe2008}), Kepler-7b ($0.35\pm0.02$; \citealt{demory2013}), WASP-12b (97.5\% confidence upper limit at 0.064; \citealt{bell2017}), WASP-18b ($<$0.048 at $2\sigma$; \citealt{shporer2019}), WASP-43b ($0.24\pm0.01$; \citealt{keating2017}), and TrES-2b ($0.014\pm0.003$; \citealt{barclay2012}). In the context of these measurements, our derived albedo for WASP-19b is relatively high, being most similar to the value for WASP-43b. Meanwhile, \citet{mallonn2019} computed self-consistent $z'$-band albedos for five hot Jupiters, including WASP-19b, and found 97.5\% confidence upper limits ranging from 0.16 to 0.38.

To assess the efficiency of day--night heat transport in the atmosphere of WASP-19b, we compare the retrieved $T$--$P$ profile and isothermal dayside temperature in Figures~\ref{fig:retrieval} and \ref{fig:isotherm} with the theoretical limiting cases for the planet's dayside temperature. The dayside equilibrium temperature can be expressed as \citep[e.g.,][]{lopez2007}
\begin{equation}\label{temps}
T_{\mathrm{eq,day}} = T_{*}\sqrt{\frac{R_{*}}{a}}\left\lbrack f (1-A_{B}) \right\rbrack^{1/4},
\end{equation}
where $A_{B}$ is the Bond albedo, and $f$ is a factor that quantifies the level of heat redistribution across the planet's surface. For the case of zero recirculation (i.e., instant reradiation), $f=2/3$, while for full recirculation (i.e., homogeneous heat redistribution), $f=1/4$. Assuming Bond albedo values between 0 and 0.3 \citep[typical for highly irradiated hot Jupiters such as WASP-19b; see, for example,][]{schwartz}, these two limiting cases correspond to temperature ranges of 2420--2650~K and 1900--2070~K, respectively. In Figure~\ref{fig:retrieval}, we have plotted the limiting equilibrium temperature values for $A_{B}=0.1$. 

The retrieved a dayside temperature of $\sim$2200~K lies in between these two extremes, suggesting moderately efficient day--night heat recirculation. This is consistent with the relatively low-amplitude \textit{Spitzer} full-orbit phase curves, as well as predictions from both three-dimensional general circulation models and one-dimensional radiative transfer models \citep{wong2016}. The contribution of the planet's thermal emission to the measured \textit{TESS}-band secondary eclipse depth is roughly 130~ppm, which means that the planet's observed brightness variation across the orbit in the \textit{TESS} bandpass is dominated by modulations in the amount of reflected light, instead of the day--night temperature contrast.

\section{Conclusions}\label{sec:con}

We have presented an analysis of the full-orbit phase curve of the WASP-19 system obtained by the \textit{TESS} mission. We measured a phase curve signal corresponding to the atmospheric brightness modulation of WASP-19b with a semiamplitude of $319\pm51$~ppm, as well as a secondary eclipse depth of $470^{+130}_{-110}$~ppm. No other significant phase curve signals were detected, consistent with theoretical predictions of Doppler boosting and ellipsoidal distortion amplitudes that are comparable to or below the measured uncertainties.

No significant phase shift in the atmospheric brightness modulation was detected, indicating that the brightest region of WASP-19b's dayside atmosphere at optical wavelengths is located near the substellar point. With a nightside flux that is consistent with zero, WASP-19b joins the growing number of highly irradiated hot Jupiters with phase curve measurements in the optical and/or infrared that show large day--night temperature contrasts --- a trend that has been predicted by numerous modeling studies \citep[e.g.,][]{cowanagol,perezbecker}.

Combining the transit and secondary eclipse depths derived from the \textit{TESS} light curve with previous measurements in the literature, we carried out retrievals of WASP-19b's secondary eclipse spectrum using the SCARLET code. We inferred that the dayside emission is consistent with an isothermal atmosphere at a temperature of $2240\pm40$~K. The secondary eclipse depths at visible wavelengths require a significant amount of reflected light from the dayside atmosphere, yielding an optical geometric albedo of $0.16\pm0.04$. Comparing the measured dayside temperature with theoretical limits on the equilibrium temperature, we find that the atmosphere of WASP-19b is moderately efficient at transporting heat from the dayside to the nightside.

Together with the other published phase curves of WASP-18 \citep{shporer2019}, WASP-121 \citep{bourrier2019,daylan2019}, and KELT-9 \citep{wong2019kelt9}, this work underscores the \textit{TESS} Mission's potential as a powerful tool for exoplanet phase curve studies. Building on the groundwork laid by similar studies in the \textit{Kepler} era, \textit{TESS} will greatly expand the possibilities of space-based time-domain science with its broad sky coverage: throughout the two-year Primary Mission and continuing through the recently approved Extended Mission, \textit{TESS} will image almost the entire sky. This exquisite coverage will enable systematic searches for phase curve signals in both known and newly discovered transiting systems, which will subsequently populate the parameter space of planets with well-constrained secondary eclipse depths and day--night temperature contrasts, and enable incisive comparisons with the trends predicted by atmospheric theory and modeling. 

One particularly fruitful avenue for follow-up study is complementary observations of secondary eclipses at thermal wavelengths, similar to those utilized in this work. Such measurements break the degeneracy between atmospheric reflectivity and thermal emission inherent in the interpretation of single-band visible-light secondary eclipse depths and allow for the derivation of the dayside geometric albedo --- an important quantity for assessing the incidence of exoplanetary clouds and their dependence on various planetary and orbital parameters. Looking to the future, finding systems with robust phase curve signals from \textit{TESS} will help identify optimal targets for intensive spectroscopic observations in transmission, emission, or across the full orbital phase with upcoming facilities such as the \textit{James Webb Space Telescope}.

\acknowledgments

Funding for the \textit{TESS} mission is provided by NASA’s Science Mission directorate. This paper includes data collected by the \textit{TESS} mission, which are publicly available from the Mikulski Archive for Space Telescopes (MAST). Resources supporting this work were provided by the NASA High-End Computing (HEC) Program through the NASA Advanced Supercomputing (NAS) Division at Ames Research Center. 
I.W.~is supported by a Heising-Simons \textit{51 Pegasi b} postdoctoral fellowship.

\software{ExoTEP \citep{benneke2019,wong2019}, BATMAN \citep{kreidberg2015}, emcee \citep{emcee}, SCARLET \citep{bennekeseager1,bennekeseager2,benneke2015,benneke2019}}


\end{document}